\documentclass{aa}
\usepackage{graphicx,natbib,psfig}
\usepackage{txfonts}
\newcommand{\ltsima} {$\; \buildrel < \over \sim \;$}
\newcommand{\gtsima} {$\; \buildrel > \over \sim \;$}
\newcommand{\lta} {\lower.5ex\hbox{\ltsima}}
\newcommand{\gta} {\lower.5ex\hbox{\gtsima}}

\newcommand{\kms}{$\rm{\,km \,s}^{-1}$}

\begin{document}

\title{Measuring supermassive black holes with gas kinematics - {\rm\bf II}.\\ 
The LINERs IC 989, NGC 5077, and NGC 6500 \thanks{Based  on observations obtained at
the  Space  Telescope Science  Institute, which is operated by the
Association of Universities for Research in Astronomy, Incorporated,
under NASA contract NAS 5-26555.}}
\author{Giovanna De Francesco
\inst{1}
\and
Alessandro Capetti
\inst{1}
\and
Alessandro Marconi \inst{2}}

\offprints{G. De Francesco}

\institute{INAF - Osservatorio Astronomico di Torino, Strada
  Osservatorio 20, I-10025 Pino Torinese, Italy\\
\email{defrancesco@oato.inaf.it,capetti@oato.inaf.it}
\and
Dipartimento di Astronomia e Scienza dello Spazio, Universit\`a di Firenze,
       Largo E. Fermi 2, I-50125 Firenze, Italy\\
\email{marconi@arcetri.astro.it}}

\date{Received / Accepted}

\abstract {We present results from a kinematical study of the gas in the
nucleus of a sample of three LINER galaxies, obtained from archival HST/STIS
long-slit spectra. We found that, while for the elliptical galaxy NGC 5077, the
observed velocity curves are consistent with gas in regular rotation around
the galaxy's center, this is not the case for the two remaining objects.
By modeling the surface brightness distribution and rotation curve from
the emission lines in NGC 5077, we found that the observed
kinematics of the circumnuclear gas can be accurately reproduced by adding to
the stellar mass component a black hole mass of $M_{\rm BH} =
6.8_{-2.8}^{+4.3}\times 10^8 M_{\odot}$ (uncertainties at a 1$\sigma$ level);
the radius of its sphere of influence ($R_{\rm sph} \sim$ 0\farcs34) is 
well-resolved at the HST resolution. The BH mass estimate in NGC 5077 is in
fairly good agreement with both the $M_{\rm BH}-M_{\rm bul}$ (with an upward
scatter of $\sim$ 0.4 dex) and $M_{\rm BH}-\sigma$ correlations (with an upward
scatter of 0.5 dex in the Tremaine et al. form and essentially
no scatter using the Ferrarese et al. form) and provides further support for 
the presence of a connection between the {\sl residuals} \rm from the $M_{\rm
BH}-\sigma$ correlation and the bulge effective radius. This 
indicates the presence of a black hole's ``fundamental plane'' in the
sense that a combination of at least $\sigma$ and $R_{\rm e}$ drives the
correlations between $M_{\rm BH}$ and host bulge properties.
  }

\titlerunning{The supermassive black hole in NGC 5077}
\maketitle

\section{Introduction}
\label{intro}

It is now clear that the presence of a supermassive black hole (SMBH)
is a common, if not universal, feature in the center of galaxies. In fact,
since active galactic nuclei (AGNs) are thought to be powered by mass
accretion onto an SMBH, the high incidence of low-luminosity AGN activity in 
nearby galaxies \citep{heckman80, maoz95, ho97a, ho97b, braatz97, barth98, 
barth99,nagar02} has led to the conclusion that a significant fraction of 
galaxies in the Local Universe must host SMBH.
This conclusion is now supported by direct measurements of SMBH masses
in the centers of nearby galaxies obtained with different techniques 
\citep[see][for a review]{ferrarese05}.
These measurements indicate that the BH mass $M_{\rm BH}$ is related to 
the properties of the host galaxy, such as bulge luminosity $L_{\rm bul}$ 
and mass $M_{\rm bul}$ \citep{kormendy95, magorrian98, marconi03}, light 
concentration \citep{graham01, graham07}, and bulge velocity dispersion 
$\sigma_{\rm bul}$ \citep{ferrarese00, gebhardt00a, tremaine02}. 

The existence of any correlations between $M_{\rm BH}$ and host
bulge properties supports the idea that the growth of SMBHs and
the formation of bulges are closely linked \citep{silk98, haehnelt00}.
This has profound implications for the process of galaxy formation and 
evolution. Moreover, SMBH mass estimates inferred via the above correlations, 
when more direct methods are not feasible, enter into a variety of important 
studies spanning from AGNs physics to the coeval formation and 
evolution of the host galaxy and its nuclear black hole.  

These results need, however, to be further investigated by increasing the 
number of accurate BH mass determinations in nearby galactic nuclei 
to set these correlations on a stronger statistical basis. 
In particular, such a study has the potential of establishing the precise
role of the various host galaxy's parameters in setting the resulting BH mass. 
To date, reliable SMBH detections have been obtained for a limited number of
galaxies ($\sim$ 30, \citealt{ferrarese05}), with the bulk of $M_{\rm BH}$ 
estimates in the range of $10^{7}-10^{9}  M_{\odot}$. 
To add reliable new points to the $M_{\rm BH}-$ host galaxy's properties
planes is then a fundamental task for future developments of astronomical 
and physical studies. 

One widely applicable and relatively simple method of detecting BHs is 
based on gas kinematics (e.g. \citealt{harms94,ferrarese96, macchetto97, 
barth01}), through
studies of emission lines from circumnuclear gas disks, provided that the gas 
velocity field is not significantly influenced by non gravitational motions. 
However, the purely gravitational kinematics of the gas can be established 
{\sl a posteriori} from successful modeling of the gas velocity field 
under the sole influence of the stellar and black hole potential.

The Space Telescope Imaging Spectrograph (STIS) onboard $\it {HST}$ is still
the most suitable instrument for such studies as it provides a high 
angular resolution ($\sim 0\farcs1$) in the optical spectral region.
In this band brighter emission lines are found with respect to the infrared, 
the only band accessible at high resolution with ground-based adaptive optics 
telescopes. The wealth of unpublished data contained in STIS archives
represents an extraordinary and still unexplored resource. With the aim of 
finding galaxy candidates to provide a successful SMBH mass measurement,  
we performed a systematic search for unpublished data in the $\it
{HST}$ STIS archive. 

In this paper we present the results obtained for a sample of three LINER
galaxies. The galaxies were observed with STIS on {$\it HST$} during Cycle 7
under Proposal ID 7354 and were part of a larger sample of eight selected
objects. The results on NGC 3998 have already been published
\citep[][hereafter Paper I]{ngc3998}. In a forthcoming paper, we will complete
the proposal targets and discuss the relation between BH mass, host galaxy 
properties, and nuclear activity. Table \ref{source} lists the galaxies and 
their main physical properties. Basic data are from the Lyon/Meudon 
Extragalactic Database (HyperLeda) or NASA/IPAC Extragalactic Database (NED).  
Distances are calculated with ${\it H}_0$ = 75 km s$^{-1}$ Mpc$^{-1}$ and 
are corrected for Local Group infall onto Virgo.

The paper is organized as follows. In Sec. \ref{obs} we present HST/STIS data
and the reduction that lead to the results described in Sec. \ref{results}. 
In Sec. \ref{fitting} we model the observed emission-line rotation curve for 
NGC 5077 and show that the dynamics of the circumnuclear gas can be 
accurately reproduced by circular motions in a thin disk when a point-like 
dark mass is added to the stellar potential.  Our results are discussed 
in Sec. \ref{discussion}, and summarized in Sec. \ref{summary}.

\section{HST data and reduction}
\label{obs}

The three galaxies considered were observed with STIS on {$\it HST$}  
with the G750M grating and the 52\arcsec $\times$ 0\farcs1 slit.
Data were acquired at different slit positions, following a
perpendicular-to-slit pattern with a step of 0\farcs1\ and the central slit
centered on the nucleus.   
The spectra obtained for each source, NUC for the nuclear slit, 
N1-N$_{\rm n}$ (to North), and S1-S$_{\rm n}$ (to South) 
for the off-nuclears, were retrieved from the public archive.  
Table \ref{source} lists the number of slit positions,  
the orientation of the slit (from north to east), the exposure time for each
positioning, and the observation date for the galaxies of the sample.  

The data were obtained with a 2x1 on-chip binning of the detector pixels 
and automatically processed through the standard {$\it CALSTIS $} pipeline 
to perform the steps of bias and dark subtraction, applying the flat field 
and combining the two sub-exposures to reject cosmic-ray events.
The data were then wavelength and flux calibrated with conversion to
heliocentric wavelengths and absolute flux units and rectified for the
geometric distortions. The 2-D spectral image obtained for each slit 
position has a spatial scale of 0\farcs0507 pixel$^{-1}$ along the slit, 
a dispersion of $\Delta \lambda$ = 1.108 \AA \ pixel$^{-1}$, and a 
spectral resolution of $\mathcal{R} \simeq 3000$, covering the rest 
frame wavelength range 6480-7050 \AA.

\begin{table*}
\caption{Sample of galaxies and STIS data.}
\begin{tabular}{l c c c c c c c}  \hline\hline 
 Galaxy &  Type & $D$ & $\sigma_{\rm star}$ & N$_{\rm slit}$  & PA & Exp. T. &  Date          \\
        &       & (Mpc) & (km s$^{-1}$)  &     &               & (s) &  \\ \hline 
IC 989  & E  & 102   &  185 & 3  &  -157  &  494   & 1998-03-13  \\       
NGC 5077  & E3-4 & 38  &  256 & 3  &  -81  &  418   & 1998-03-12    \\            
NGC 6500  & Sab   &  42 &  200  & 5  & 15  &  236  & 1998-11-03    \\       
\hline
\end{tabular}
\label{source}
\end{table*}

For each spectrum we selected the regions containing the lines of interest. 
The lines were fitted, row by row, along the dispersion direction, together 
with a linear continuum, with Gaussian functions using the task SPECFIT in 
STSDAS/IRAF. All emission lines present in the spectra (H$\alpha$,
[N II]$\lambda\lambda$6548,6583 and [S II]$\lambda\lambda$6716,6731) were
fitted simultaneously with the same velocity and width and with the relative 
flux of the [N II] lines kept fixed to 0.334. 
The assumption relative to the sharing of the same kinematics for the emission 
lines will result in average kinematical quantities, weighted with the line 
surface brightness\footnote{However, we investigated the possibility that 
lines did not share common kinematics, but we did not find evidence of this.}. 

\begin{figure}
\centering{
\psfig{figure=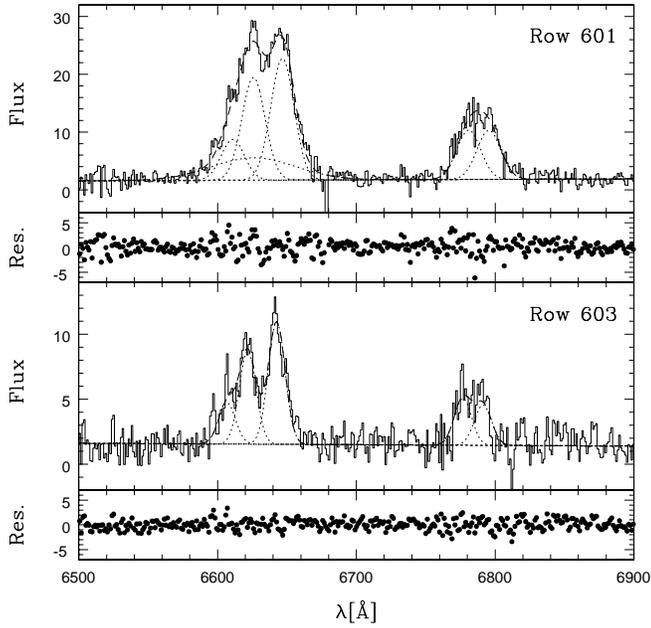,width=1.0\linewidth}}
\caption{\label{data} Nuclear spectra of NGC 5077: at the slit center position 
({\it upper panels}), showing the possible presence of a weak broad H$\alpha$ 
component, and at 0\farcs1 from it ({\it lower panels}). 
The best model spectrum (dashed line) is superposed on the data; dotted 
lines are for the Gaussians and linear continuum components of the fit.
Flux is in units of 10$^{-15}$ erg cm$^{-2}$ s$^{-1}$ \AA$^{-1}$ arcsec$^{-2}$. 
The residuals from the fits are shown in the smaller panels.} 
\vskip 0.5cm
\end{figure}

In the regions where the signal-to-noise ratio (SNR) was insufficient, the 
fitting was improved by co-adding two or more pixels along the slit direction. 
As an example of the quality of data, Fig. \ref{data} shows the spectrum of 
NGC 5077, the only galaxy of the group with a marginal detection of a broad 
H$\alpha$ component, at two different positions along NUC slit. 

\section{Results}
\label{results}

The results obtained from the fitting procedure for the three galaxies 
are shown in Figs. \ref{group1} through \ref{group2}, where we plot the
kinematical quantities, central velocity, and velocity dispersion, together 
with the surface brightness of the emission line with the best SNR at each 
location along the slit. The position-velocity diagram for the off-nuclear 
with the best quality of data is also reproduced in the upper panels of 
Figs. \ref{group1} and \ref{group2} for two sources of the program. 
In the following we discuss the results obtained for each galaxy of the sample. 

\subsection{IC 989}
\label{IC 989}

\begin{figure}
\centering{
\psfig{figure=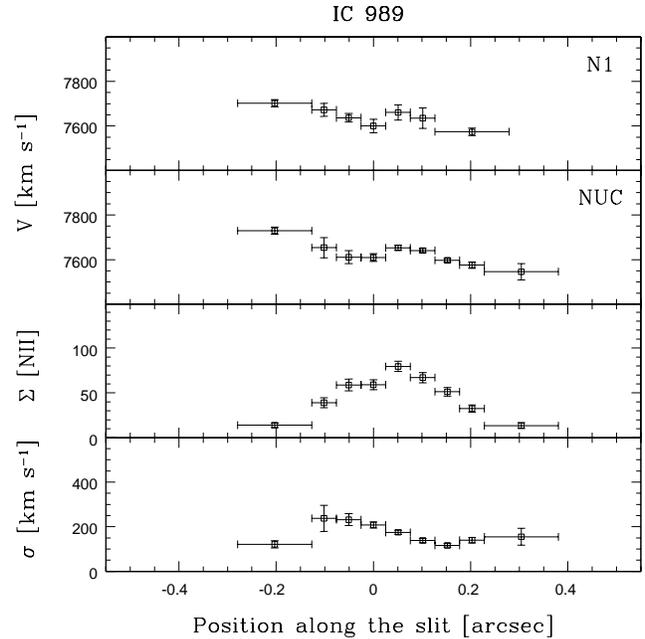,width=1.0\linewidth}}
\caption{\label{group1} {\it Upper panel}: Velocity curve from the off-nuclear 
slit with the best quality of data. {\it Lower panels}: Velocity, surface 
brightness, and velocity dispersion along NUC slit. Surface  brightness is in 
units of 10$^{-15}$ erg s$^{-1}$ cm$^{-2}$ arcsec$^{-2}$. Positions along the
slit are relative to the continuum peak.} 
\vskip 0.5cm
\end{figure}

The velocity curve on the nuclear slit shows a large-scale rotation with 
redshift at the negative space values (NE region) and blueshift at the
positive (SW region) (Fig. \ref{group1}).  
The same trend is observable in the off-nuclear slits, with almost the same 
values of velocity gradients. However, the velocity gradient is reversed at 
the very center, suggesting the presence of a counter-rotating nuclear 
gas disk.   

\subsection{NGC 5077}
\label{NGC 5077}

\begin{figure*}
\centerline{
\psfig{figure=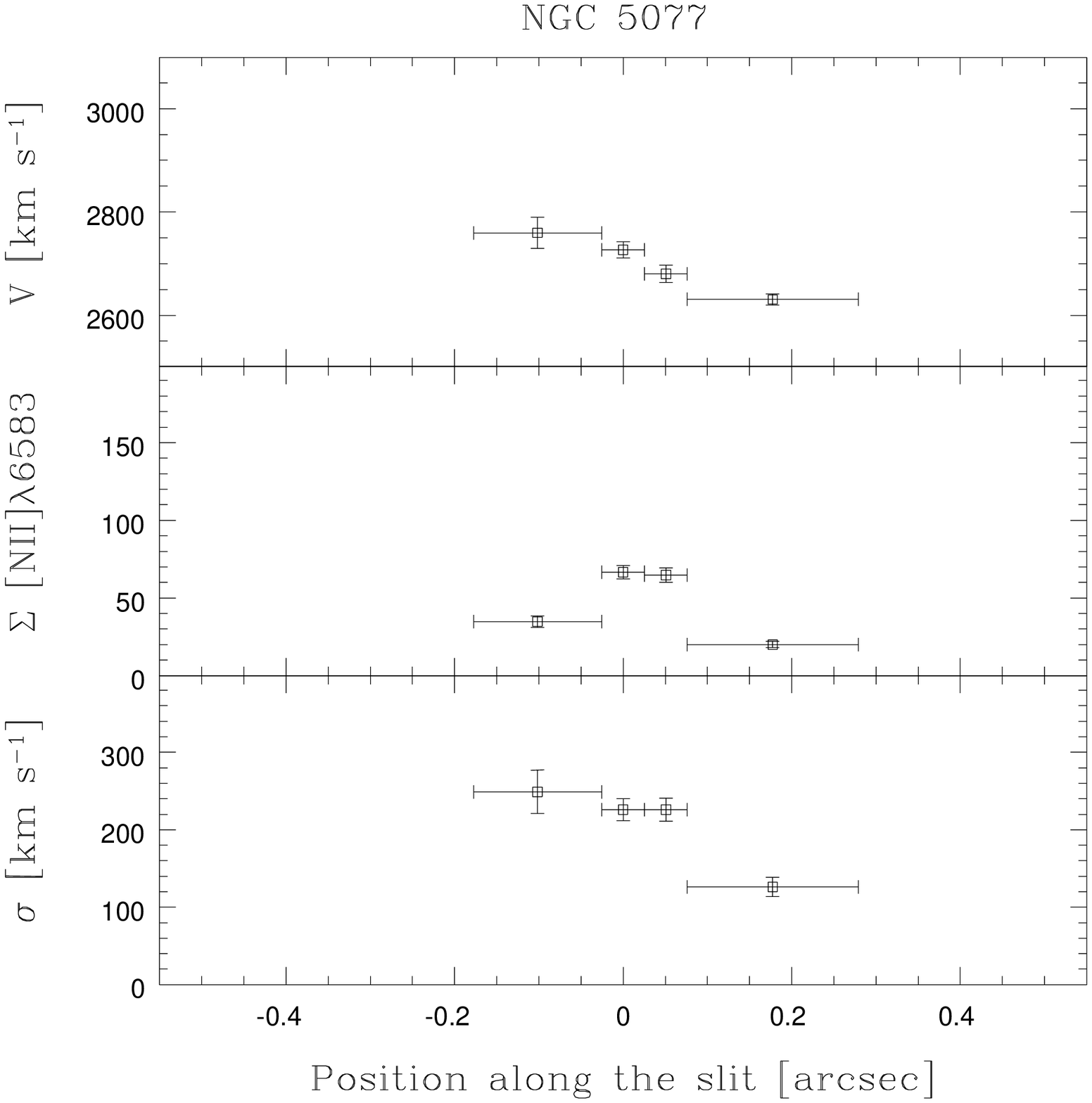,width=0.33\linewidth}
\psfig{figure=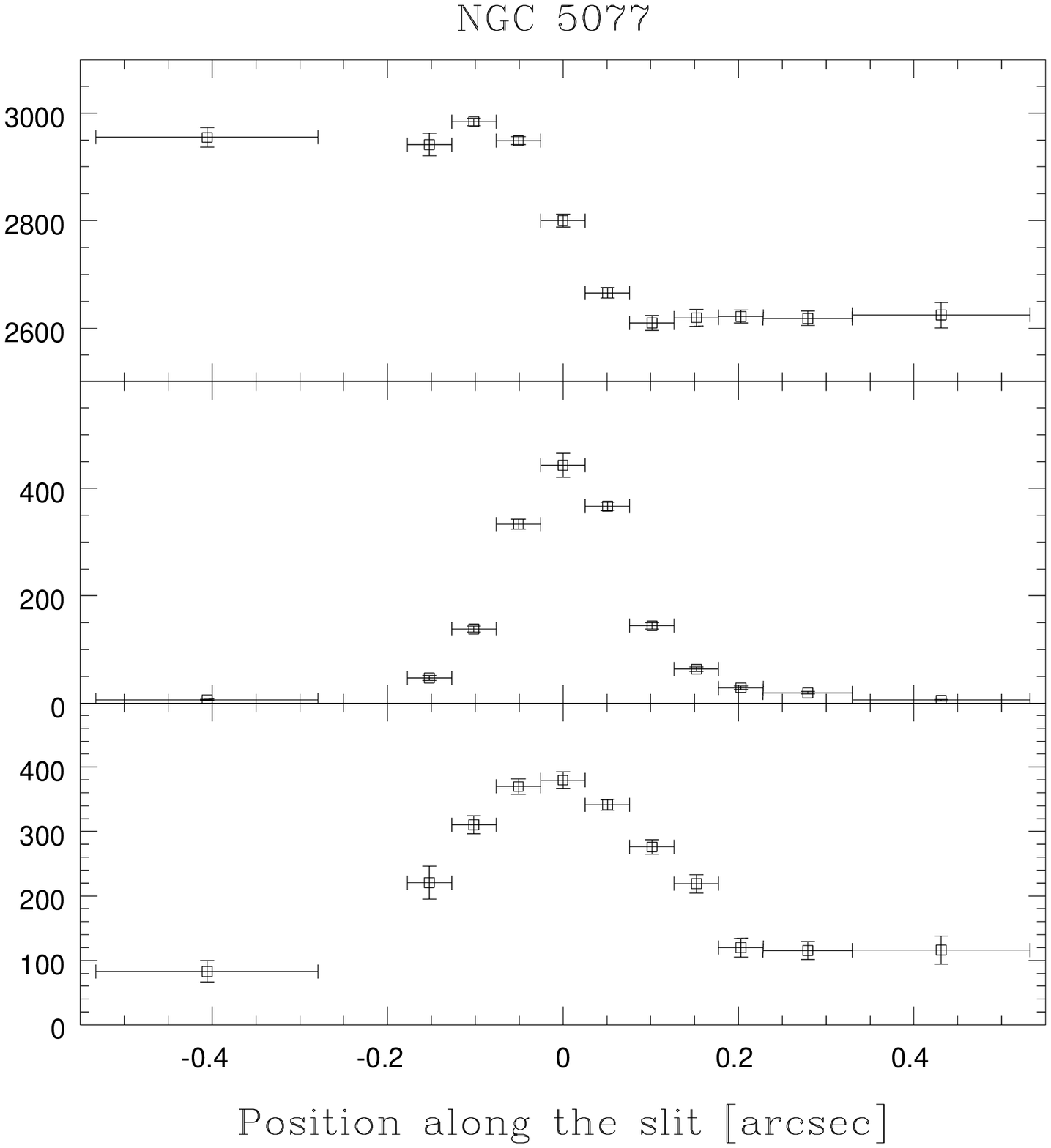,width=0.33\linewidth}
\psfig{figure=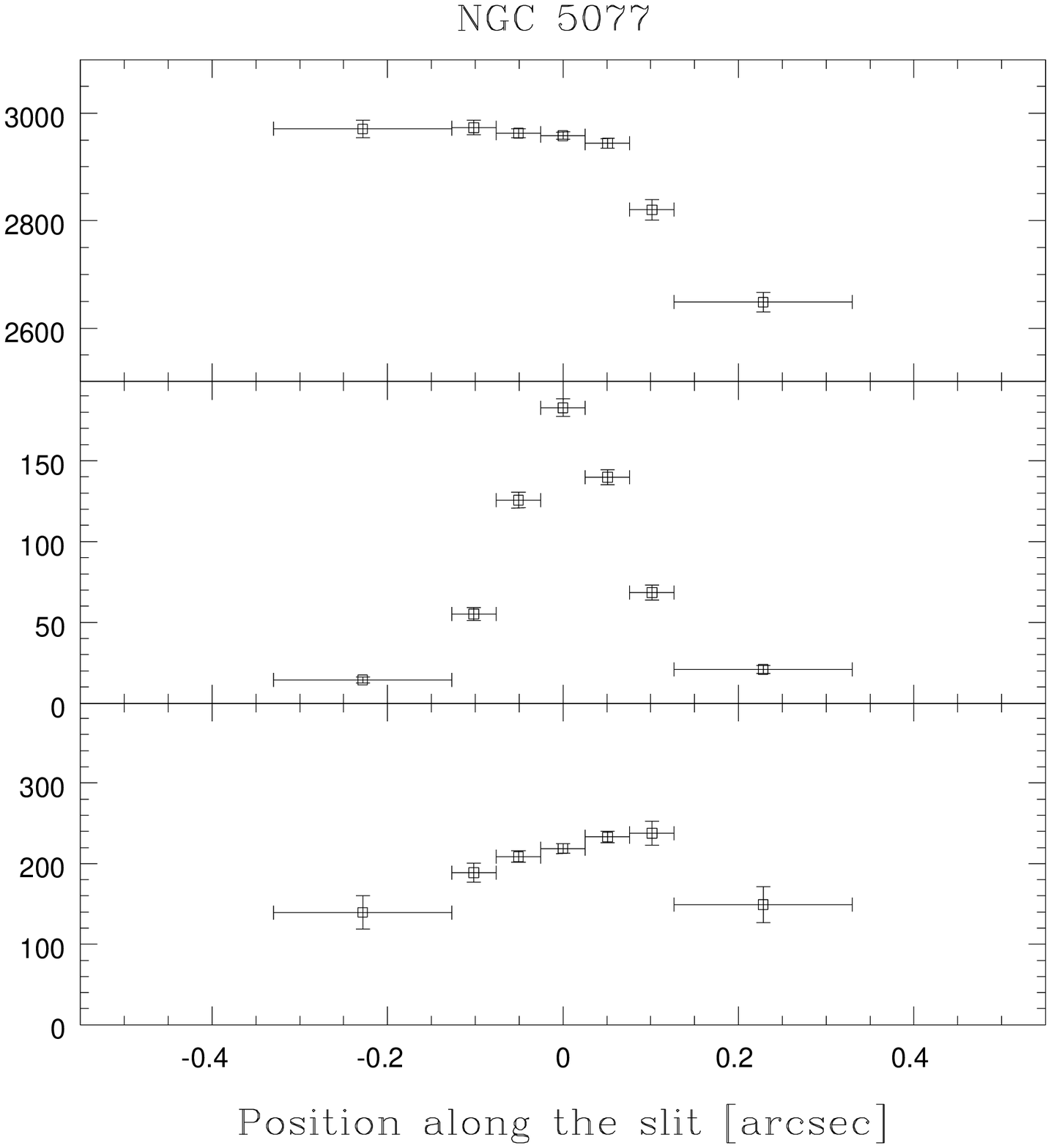,width=0.33\linewidth}}
\caption{\label{far} Velocity, surface brightness, and velocity dispersion 
along the off-nuclear slit position N1 ({\it left panel}), NUC ({\it central 
panel}), and S1 ({\it right}). Positions along the slits are relative to the 
continuum peak, positive values are NW.} 
\vskip 0.5cm
\end{figure*}

For the elliptical galaxy NGC 5077, the observed velocity curves are
apparently consistent with gas in regular rotation around the galaxy's center. 
The velocity curve in the central slit, NUC (Fig. \ref{far}, central panel),  
has a full amplitude of $\sim 400$ \kms and shows a general reflection
symmetry: starting from the center, the velocity rises rapidly on both sides 
by $\sim 200$ \kms reaching a peak at $r \sim$ 0\farcs1 from the center. 
This trend is followed by a small decrease and a substantially constant value 
at larger radii. Line emission and velocity dispersion are strongly peaked and
smoothly decrease from the nucleus outwards.

The behavior seen in the off-nuclear slits is qualitatively similar
to what is seen at the NUC location (Fig. \ref{far}, left and right panels), 
but with smaller velocity amplitude and, more important, a less extreme
velocity gradient. Both the amplitude and gradient decrease at increasing
distance of the slit center from the nucleus, with a behavior characteristic 
of gas rotating in a circumnuclear disk. 

Ground-based photometric and kinematical studies \citep{demoulin84, bertola91} 
show that NGC 5077 exhibits a gaseous disk with the major axis roughly 
orthogonal to the galaxy photometric major axis (PA $\sim 10^\circ$). 
The gas isophotes show twisting and a marked warp on the W side at $r \sim$
20\arcsec \citep{caon00}.  
The gas has a fairly symmetric and smooth rotation curve at PA = $108^\circ$ 
($\sim$ major axis of the ionized gas distribution) with a half amplitude of 
$\sim 270$ \kms at $r \sim$ 13\arcsec. At PA = $10^\circ$, the stellar
rotation curve exhibits a counter-rotating core ($r \leq$ 5\arcsec). Along 
this axis, the gas rotates in the same direction as the stellar nucleus and 
shows a small-scale central velocity plateau.       

\subsection{NGC 6500}
\label{NGC 6500}

\begin{figure}
\centering{
\psfig{figure=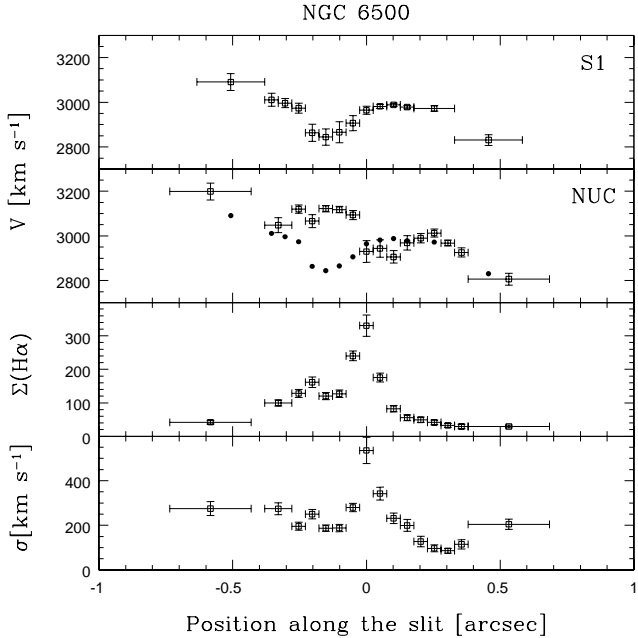,width=1.0\linewidth}}
\caption{\label{group2} Same as Fig.\ref{group1}. {\it Second upper panel}: 
Over-plot of NUC ({\it empty squares}) and S1 ({\it filled circles}) velocity 
curves showing the inconsistency of the gas position-velocity diagram with 
regular rotation of a nuclear gas disk.}  
\vskip 0.5cm
\end{figure}

The velocity curves for this galaxy are complex (see Fig. \ref{group2}).  
On the nuclear slit, the velocity apparently shows a general rotational trend, 
with redshift on the left and blueshift on the right side of the diagram.  
The same large-scale rotation is observable on the off-nuclear slit
positions. Smaller scale high-amplitude ($\sim$ 200 \kms) velocity 
oscillations are also clearly visible.    

In the SW region at $r \leq$ 0\farcs4 the position-velocity diagrams of 
NUC and the adjacent S1 slit show a trend that is inconsistent with gas 
in regular rotation around the galaxy's center (see Fig. \ref{group2}, 
second upper panel). In this region the velocity curves diverge, reaching a
separation of $\sim$ 300 \kms at $r \sim$ 0\farcs2. The over-plot of the 
position velocity diagrams suggests an expanding bubble of gas .  

\begin{figure*}
\hskip 0.25cm
\psfig{figure=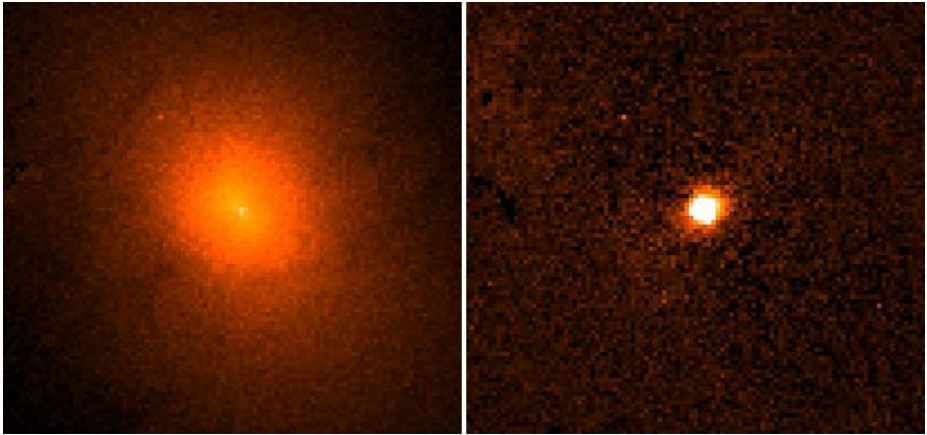,width=0.68\linewidth,angle=0}
\centering
\vskip 0.5cm
\caption{\label{vima} {\it Left}: {\it V} band ({\it F547M} filter) image of 
NGC 5077. {\it Right}: H$\alpha$ WFPC2/HST image. Field of view 
4\farcs5 $\times$ 4\farcs5. North is up, East is left.}
\end{figure*}

Summarizing, we find that NGC 5077 shows a line emission consistent with a 
circumnuclear gas disk in Keplerian rotation around the galaxy's center.   
For the remaining two objects, the trend of their velocity curves cannot be 
ascribed to the regular rotation of a nuclear gas disk. The position-velocity 
diagrams of IC 989 indicate the presence of a counter-rotating gas component. 
Modeling this configuration would require (at least) a warped geometry for 
the disk, which cannot be constrained with the present data. By these
considerations, only NGC 5077 is a suitable candidate for providing a 
successful SMBH mass measurement. 

\section{Modeling the rotation curves of NGC 5077}
\label{fitting}

Our modeling code, described in detail in \citet{marconi03_2}, was used to fit 
the observed rotation curves of NGC 5077. The code computes the rotation
curves of the gas assuming that the gas is rotating in circular orbits within 
a thin disk in the galaxy potential. The gravitational potential has two 
components: the stellar potential (whose mass distribution will be determined 
in Sec. \ref{stelle}), characterized by its mass-to-light ratio and a dark
mass concentration (the black hole), spatially unresolved at HST+STIS
resolution and characterized by its total mass $M_{\rm BH}$.
In computing the rotation curves, we take into account the finite spatial 
resolution of the observations, and the intrinsic surface brightness 
distribution (ISBD) of the emission lines and we integrate over the slit 
and pixel area. The adopted HST+STIS point spread function (PSF) is obtained 
by fitting with three Gaussians the Tiny Tim \citep{krist99} model PSF 
calculated at 6700 \AA without the Lyot-stop, according to the procedure 
suggested by \citet{dressel06}. The $\chi^2$ is minimized to determine the 
free parameters using the downhill simplex algorithm by \citet{press92}.

\subsection{The stellar mass distribution}
\label{stelle}
\begin{figure}
\centering
\psfig{figure=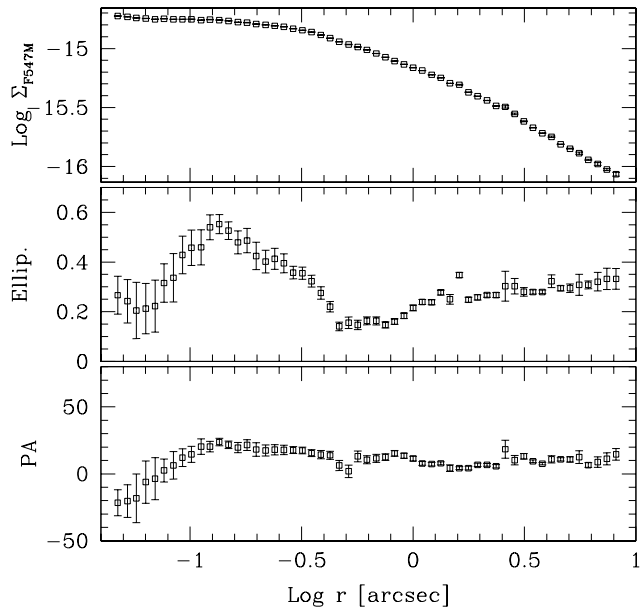,width=1.0\linewidth,angle=0}
\vskip 0.5cm
\caption{\label{ellipse} Results of the isophotes analysis of the V band
image of NGC 5077. Surface brightness is shown in the top panel
(in units of erg s$^{-1}$ cm$^{-2}$ \AA$^{-1}$ arcsec$^{-2}$), and the
galaxy's ellipticity and position angle are shown in the middle and
bottom panels, respectively.}
\end{figure}

To assess the contribution of stars to the gravitational potential
in the nuclear region, we derived the stellar luminosity density from the 
observed surface brightness distribution. We reconstructed the galaxy light 
profile using a WFPC2 F547M ({\it V} band) image retrieved from the public 
archive (Fig. \ref{vima}). 
Based on WFPC2 F702W ({\it R} band) images, \citet{tran01} found a filamentary
dust structure in NGC 5077, with position angle of the major axis of the main
dust feature $\sim 102^\circ$ (at $r$ = 10\arcsec). Our inspection of {$\it HST$} 
archive images, however, shows that the influence of dust is negligible. 
In fact, \citet{tran01} derived a mean visual extinction of only $<A_V>$ = 0.048. 
We used the IRAF/STSDAS program ELLIPSE to fit elliptical isophotes to the galaxy 
(see Fig. \ref{ellipse}). Nuclear regions ($r \leq$ 0\farcs3) show the largest 
ellipticity variations. At larger radii, ellipticity and position angle show
small variations around the values 0.25 and 10$^\circ$, respectively. 

\begin{figure}
\centering
\psfig{figure=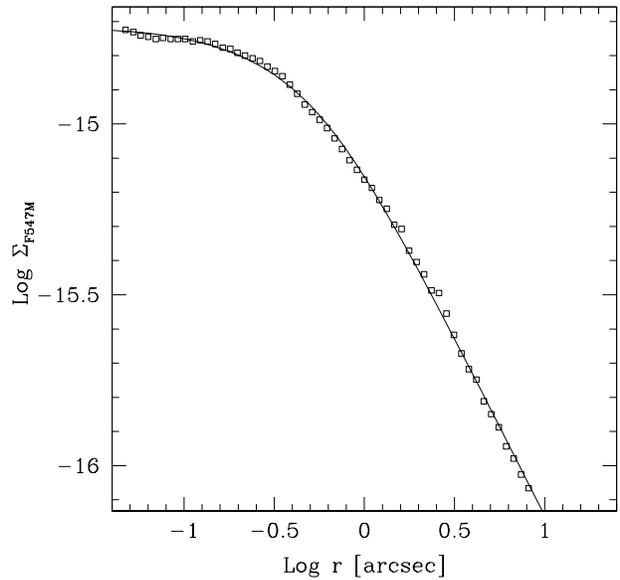,width=1.0\linewidth,angle=0}
\vskip 0.5cm
\caption{\label{mass} Fit to the surface brightness profile obtained from 
an oblate spheroid stellar density distribution.}
\end{figure}

We derived the stars' density profile from the galaxy surface brightness
following the same method described in Paper I when assuming an oblate 
spheroid density distribution. Following \citet{marel98}, the stellar density
distribution was parameterized as 
$$ \rho(m) = \rho_0\left(\frac{m}{r_{\rm b}}\right)^{-\alpha}
\left[1+\left(\frac{m}{r_{\rm b}}\right)^2\right]^{-\beta}
$$ where $m$ is given by $m^2 = x^2+y^2+z^2/q^2$, $xyz$ is a reference
system with the $xy$ plane corresponding to the principal plane of the
potential and $q$ is the intrinsic axial ratio. The above stellar density 
is integrated along the line of sight, to provide the surface brightness 
distribution on the plane of the sky, and is then convolved with the 
telescope+instrument PSF (derived from Tiny Tim) and averaged over the
detector pixel size to obtain the observed surface brightness distribution 
\citep[for details see][]{marconi03_2}. The de-projection was performed 
by adopting the reference value of ${\it\Upsilon_V}$ = 1 for the 
mass-to-light ratio in the V band (its real value will be determined in 
Sec. \ref{expl}). As the intrinsic potential axial ratio $q$ is related to 
the inclination of the principal plane with respect to the line of sight 
by the observed isophote ellipticity, we are left with the freedom of 
assuming different galaxy inclinations to the line of sight. We performed 
the de-projection adopting the value $i = 44^\circ$ for the galaxy
inclination, as given from the HyperLeda database. Due to the small 
ellipticity of this galaxy, the precise value of its inclination only has a 
marginal effect on the resulting mass distribution. The best fit obtained is 
shown in Fig. \ref{mass} with $\alpha$ = 0.47, $\beta$ = 0.79, and 
$r_{\rm b}$ = 0.59\arcsec.

\subsection{Fitting the gas kinematics}
\label{expl}

\begin{figure*}
\centerline{
\psfig{figure=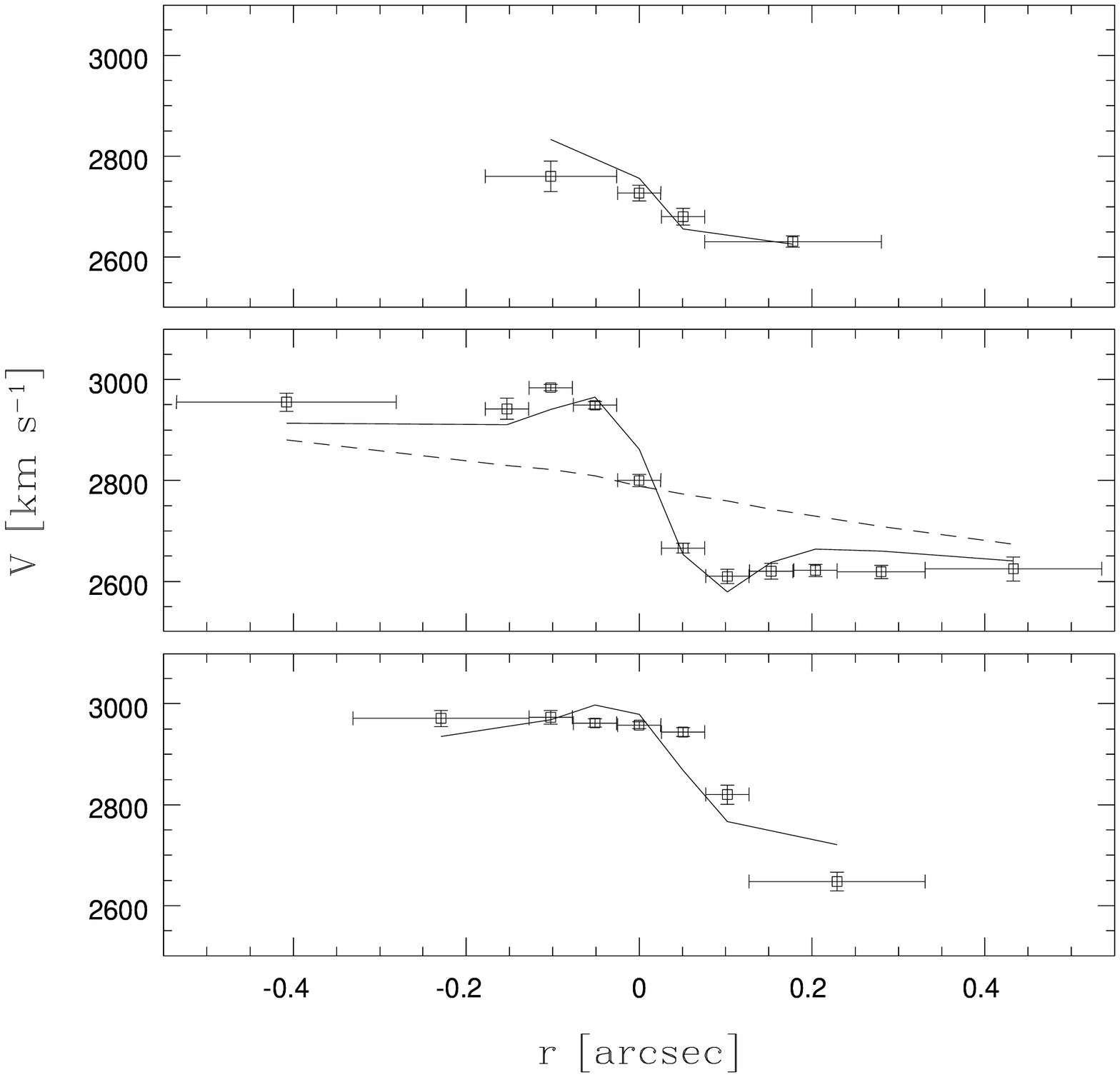,width=0.5\linewidth}
\psfig{figure=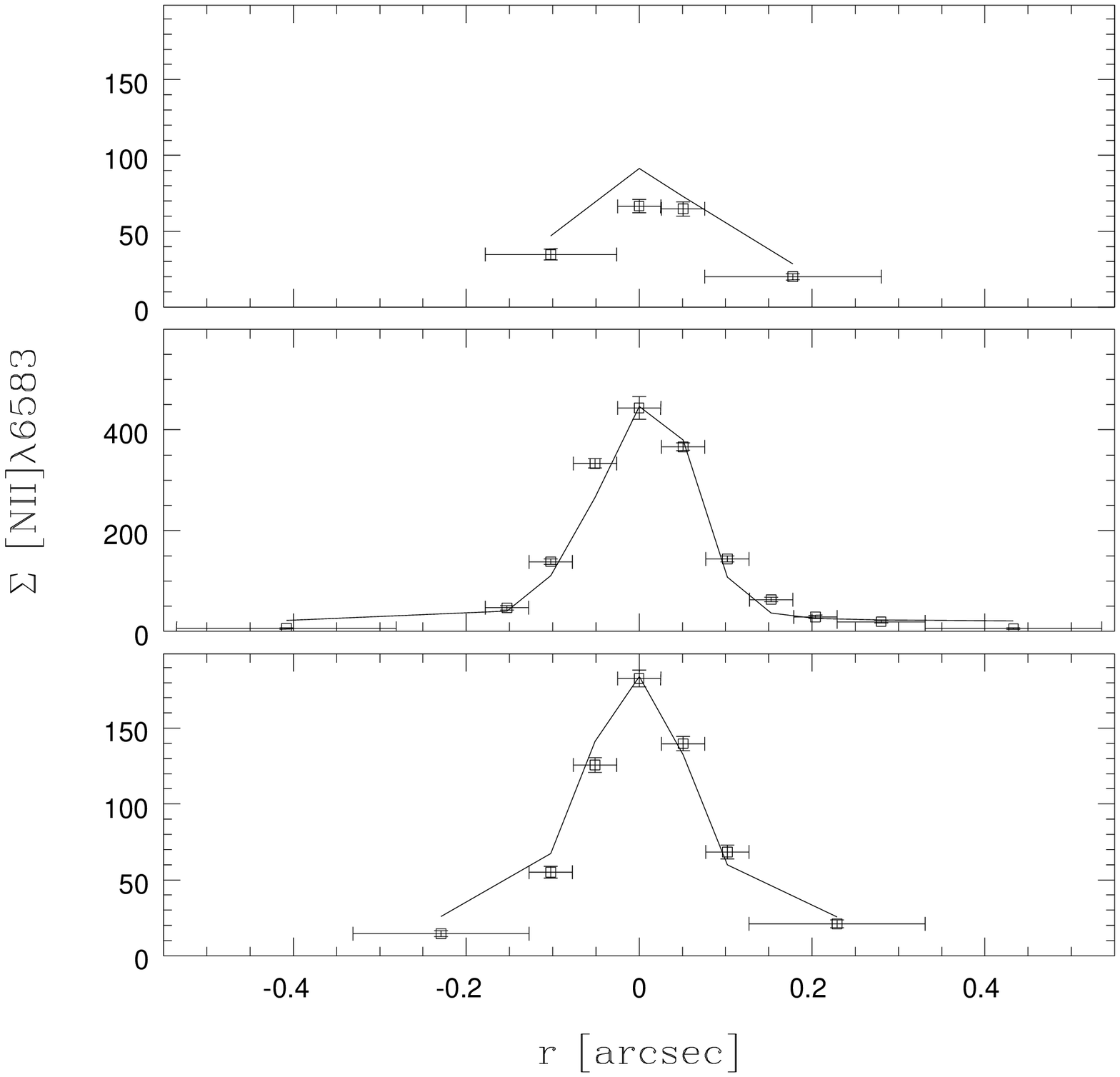,width=0.5\linewidth}}
\caption{\label{fit30} Overall best fit to the rotation curves ({\it left})  
with the case of null BH mass (dashed line) superposed on the central slit. 
Corresponding fit to the line surface brightness distribution ({\it right}). 
From upper to bottom panel: N1, NUC, S1.}
\end{figure*}

The observed kinematical quantities are averages over apertures 
defined by the slit width and the detector pixel size along the slit. 
The model fitting to the observed kinematical quantities thus depends  
on the intrinsic emission-line surface-brightness distribution, which 
is the weight for the averaging process, and on the following parameters: 
\begin{itemize}
\item
the black hole mass $M_{\rm BH}$;
\item
the mass-to-light ratio of the stellar component, ${\it\Upsilon_V}$;
\item
the position of the kinematical center in the plane of the sky, $x_0$, $y_0$, 
with the reference frame centered on the nuclear slit and $x$ axis aligned 
with the slit; 
\item
the inclination of the rotating disk, $i$;
\item
the angle between the slits and the line of nodes, $\theta$;
\item
the systemic velocity of the disk, $v_{\rm sys}$.
\end{itemize} 

With respect to Paper I, the modeling code has been slightly improved, 
so it now allows us to 1) fit the surface brightness and kinematical data  
simultaneously and 2) set limits to the parameters range. We used the second 
feature in particular to constrain the mass-to-light ratio to 
${\it\Upsilon_V} < 7.87$, as derived from \citet{maraston98} for a Salpeter 
initial mass function (IMF). 

A crucial issue for modeling the gas kinematics is the inclination of the 
nuclear gas disk, as this is strongly coupled with the black hole mass and 
${\it\Upsilon_V}$. In an oblate spheroid, the stable orbits of the gas are 
coplanar with the principal plane of the potential, and it is possible to 
directly associate the galaxy inclination and line of nodes with those of 
the circumnuclear gas. However, the potential shape is not determined well 
enough by the isophotal fitting down to the innermost regions of 
the galaxy, and it is possible that a change of principal plane might occur 
at the smallest radii, in particular within the sphere of influence of a 
supermassive black hole. We then performed a $\chi^2$ minimization for 
different values of $i$, namely $i= 20^\circ,..., 70^\circ$, allowing 
all other parameters to vary freely, including those describing the 
intrinsic line surface brightness distribution. Due to the sensitivity of 
the velocity dispersions to the ISBD modeling and to other computational 
problems (i.e. a coarse sampling), as shown by \citet{marconi06}, we 
decided to initially restrict the kinematical fitting only to the velocity 
curves. We show in Sec. \ref{vdisp} that including the velocity
dispersions in the fitting procedure only has a marginal effect on our
results.  

Inspection of Fig. \ref{far} suggests that the ISBD should be described at 
least by 2 components: one at the NUC slit (the nuclear component), 
plus a more extended one (base component). The intrinsic emission line 
surface brightness was modeled with a composition of two circularly symmetric 
Gaussians and a constant function (having as free parameters the amplitudes, 
scale radii, and the positions on the sky plane of the symmetry centers). 
\citet{marconi06} have recently shown that the choice of a particular model 
for the line surface brightness distribution affects the final BH mass 
estimate very little, provided that the model reproduces the observed line 
emission within the errors. They showed that systematic errors on $M_{\rm BH}$ 
due to the adopted intrinsic emission line surface brightness were on the 
order of 0.05 in log $M_{\rm BH}$ for their STIS data. We also tested the 
effect, for our data, of using exponential functions in the ISBD modeling. The 
BH masses derived were $\la $ 0.03 in log $M_{\rm BH}$ lower than obtained 
using a Gaussian ISBD.       
 
From the best fits obtained at varying gas disk inclination, we quoted the
goodness of the fits only with respect to the velocity curves, since we are
interested in the kinematical model that reproduces the position-velocity data
best. Excluding the contribution to the $\chi^2$ of the ISBD fit is a
conservative choice for our data, as we verified that quoting the global fits
would narrow the ranges of acceptable model parameters. The overall
best-fitting model, presented in Fig. \ref{fit30}, is obtained for $i=
40^\circ$, $M_{\rm BH} = 6.8\times 10^8 M_{\odot}$, ${\it\Upsilon_V}$ = 7.7,
and an angle from the slits to the line of nodes of 34$^\circ$ (i.e. offset by
25$^\circ$ from the ground-based measurements).  Table \ref{bestfit} shows the
best kinematical model parameters, together with the corresponding $\chi^2$
reduced value ($\chi^2_{\rm r}$ = $\chi^2$/d.o.f. with d.o.f. = 16 degrees of
freedom).  Apparently, the best fit slightly underpredicts the large-scale
velocity points. We then tested the effect of allowing a higher maximum value
of the mass-to-light ratio, related to a different choice of the IMF.  A fit
was performed by adopting the value 13.78 as limit to ${\it\Upsilon_V}$,
corresponding to the oldest stellar population and a giant-dominated IMF.
Indeed, the best fit obtained (with ${\it\Upsilon_V} = 13.2$) reproduces the
observed velocities better at large radii. However, the BH mass does not
change significantly, resulting in log $M_{\rm BH}$ only 0.02 lower than our
previous result. We therefore preferred to maintain the constraint on
${\it\Upsilon_V}$ derived by adopting the standard Salpeter IMF.

\begin{figure}
\centerline {
\psfig{figure=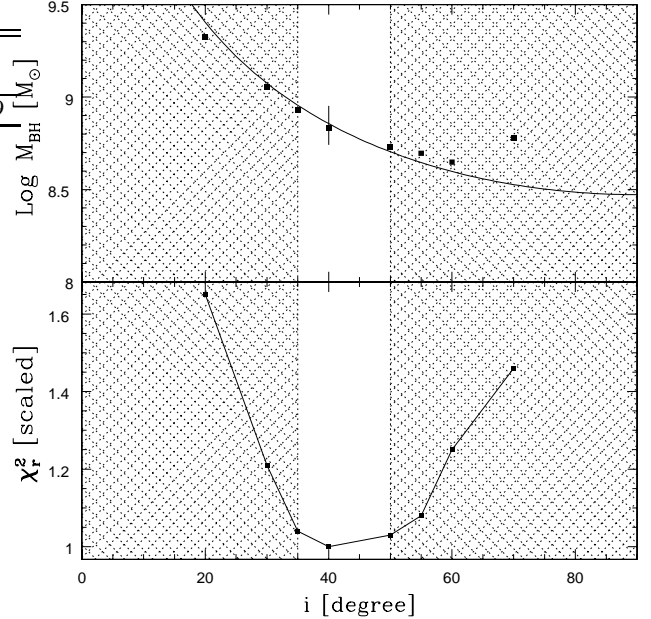,width=1.0\linewidth}}
\caption{\label{gamma} {\it Upper panel}: Best fit values of $M_{\rm BH}$
obtained by varying the disk inclination. The continuous line reproduces the
$\propto$ 1/$\sin ^2 i$ dependence of $M_{\rm BH}$. {\it Lower panel}: 
Corresponding $\chi^2_{\rm r}$ values scaled to the overall best fit.   
Inclinations smaller than 35$^\circ$ or larger than 50$^\circ$ are excluded 
at a confidence level of 1$\sigma$ (shaded regions in the diagrams), 
evaluated for 16 degrees of freedom.}
\end{figure}

The value of minimum $\chi^2_{\rm r}$ (see Table \ref{bestfit}) is far 
higher than the value indicative of a good fit. To establish the 
statistical significance of $\chi^2_{\rm r}$ variations among different 
models, we followed \citet{barth01} and rescaled the error bars in our 
velocity measurements by adding a constant error in quadrature such that 
the overall best-fitting model provides $\chi^2_{\rm r} \sim$ 1. This is 
quite a conservative approach as it has the effect of increasing the final 
uncertainty on $M_{\rm BH}$. The additional velocity error is 45 km s$^{-1}$. 
The rescaled $\chi^2_{\rm r}$ values against inclination are reported in 
the bottom panel\footnote{The different trend (high increase) 
of $\chi^2_{\rm r}[scaled]$ at small angles with respect to NGC 3998 
(Paper I, Fig. 11) is determined by the a priori fixing of an upper limit 
for ${\it\Upsilon_V}$ values.} of Fig. \ref{gamma}. The acceptable models 
(within the 1$\sigma$ confidence level, 
i.e. $ \Delta \chi^2_{\rm r} \leq 0.06$ for 16 d.o.f.),  
correspond to the range of inclination of $35^\circ \leq i \leq 50^\circ$. 

The case of no black hole mass is also shown in Fig. \ref{fit30} for the
central slit. This model (obtained by constraining ${\it\Upsilon_V} < 7.87$ and
$v_{\rm sys}$ to the same value of the best fit) predicts a velocity gradient 
that is too shallow with respect to the data, particularly in the central 
region $r \leq 0\farcs2$, with a correspondingly unacceptable value of 
$\chi^2_{\rm r}$ = 7.9 (properly rescaled to the overall best fit).  

\begin{table}
\caption{Overall best-fit parameters.}
\begin{tabular}{c c c c c c c c} \hline\hline
$i$ & $x_0$ & $ y_0$ & $\theta$ & $ v_{sys}$ & ${\it\Upsilon_V} $ & $M_{\rm
    BH}$ & $\chi^2_{\rm r}$  \\
$({^\circ})$ &  (\arcsec) &  (\arcsec) &  $({^\circ})$ &  (\kms) &
  $(M_{\odot}/L_{\odot})_V$ & $(M_{\odot}$) & \cr  \hline
40 & 0.03 & -0.01 & 34 &
2780 & 7.7 & 6.8$\times10^8$ & 15.49 \\ \hline
\end{tabular}
\label{bestfit}
\end{table}

\begin{figure}
\centerline{
\psfig{figure=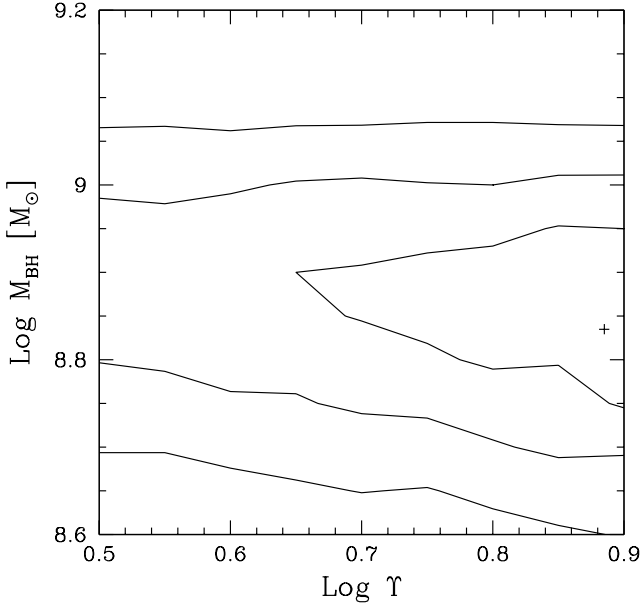,width=1.0\linewidth,angle=0}}
\caption{\label{grid} $\chi^2_{\rm r}$ contours at varying
${\it\Upsilon_V}$ and  M$_{\rm BH}$ at an inclination of $i = 40^\circ$. 
Contours are plotted for confidence levels of 1, 2, and 3$\sigma$. 
The plus sign marks the overall best fit.} 
\end{figure}

To evaluate the uncertainty associated to the black hole mass estimate, we
explored its variation with respect to the parameter that is more strongly
coupled to it, i.e. the mass-to-light ratio ${\it\Upsilon_V}$ (having already
considered the dependence on orientation).  The uncertainty on $M_{\rm BH}$
associated to changes in ${\it\Upsilon_V}$ has been estimated by building a 
$\chi^2_{\rm r}$ grid in the $M_{\rm BH}$ vs. ${\it\Upsilon_V}$ parameter
space. At each point of the grid, described by a fixed pair of $M_{\rm BH}$
and ${\it\Upsilon_V}$ values, we obtained the best fit model allowing all
other parameters to vary freely and derived the corresponding $\chi^2_{\rm r}$ 
value (properly rescaled).  The result of this analysis at $i = 40^\circ$ is
presented in Fig. \ref{grid}. The 1$\sigma$ range of the black hole mass at
this disk inclination is $M_{\rm BH} = 6.8_{-1.3}^{+2.1}\times 10^8 M_{\odot}$, 
reported in the error bar in Fig. \ref{gamma}.

If we conservatively adopt the same fractional uncertainties on $M_{\rm BH}$,
associated to variations in ${\it\Upsilon_V}$, for the allowed disk 
inclinations, we obtain a global range of acceptable black hole mass of
$M_{\rm BH} = (4.0 - 11.1)\times 10^8 M_{\odot}$ at a 1$\sigma$ level.

\subsubsection{Velocity dispersion distribution}
\label{vdisp}

At this point in our analysis we tested the influence on the above
results of including the velocity dispersions in the fitting procedure. 
The observed ionised-gas velocity dispersions are fairly large in
NGC 5077, reaching almost 400 \kms in the nuclear slit. 
Velocity dispersions larger than expected from unresolved rotation 
could be an indication of non circular motions that could invalidate 
the BH mass estimate \citep{barth01, verdoes00, verdoes02, cappellari02}. 
However, in the case of NGC 5077, rotational and instrumental broadening 
is sufficient to reproduce the behaviour of velocity dispersions.  
 
\begin{figure}
\centering
\psfig{figure=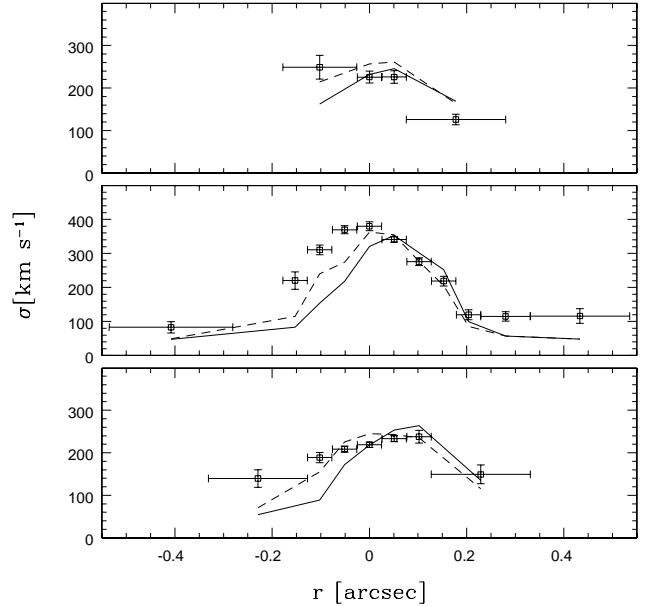,width=1.0\linewidth,angle=0}
\caption{\label{fw} Velocity dispersion distribution expected from the 
overall best fit model (solid line) at $i = 40^\circ$, and derived (dashed 
line) by including the observed velocity dispersions in the fitting
procedure. From upper to bottom panel: N1, NUC, S1.}
\end{figure}

First of all, by adopting the best fit parameter set (Table \ref{bestfit}), 
we obtained the velocity dispersion distribution shown in Fig.\ref{fw}. 
The observed velocity dispersions are reproduced acceptably well by the 
model, with only a small mismatch of observed and model
peak positions. Thus the nuclear rise of the velocity dispersion is 
accounted for as unresolved rotation by the fitting model.    

Furthermore, we repeated the $\chi^2$ minimization at $i = 40^\circ$, this 
time by including the observed velocity dispersions in the fitting procedure 
(with no free parameters added). The best fit obtained is shown in
Fig. \ref{fw}. An improvement of the match with observed values 
is obtained, while there is a slight worsening of the fit to the velocity 
curves, with $\chi^2_{\rm r}$ (scaled to the overall best fit) = 1.08.  
The effect of including the line widths in the analysis results in 
increasing the BH mass by only $\sim$ 24\% to $M_{\rm BH} = 
8.4\times 10^8  M_{\odot}$. 

\section{Discussion}
\label{discussion}

The model fitting performed for NGC 5077 indicates that the kinematics 
of gas in its innermost regions is described correctly as circular motions 
in a thin disk when a point-like dark mass of $M_{\rm BH} = 
6.8_{-2.8}^{+4.3}\times 10^8 M_{\odot}$ is added to the galaxy potential. 
We can now explore how BH mass relates to the properties of the host 
galaxy, including spheroid (bulge) mass\citep{marconi03}, stellar velocity 
dispersion \citep{tremaine02,ferrarese05}, and light concentration 
(Graham \& Driver 2007). 

Following \citet{marconi03}, we used the virial mass $M_{\rm vir}$ = $k R_{\rm
e}\sigma_{\rm bul}^{2}/G$, with $k$ = 5 \citep{cappellari06} to determine the
bulge mass of NGC 5077. We found discrepant values for the effective radius 
of the bulge, $R_{\rm e}$, in the literature. The values were derived through 
different methods (i.e. isophotal fitting, photometric aperture growth curves) 
and are 15\farcs4 \citep{bender89}, 15\farcs9 \citep{sanchez04}, 17\farcs3 
\citep{poulain94}, 19\arcsec \citep{bertola91}, 22\farcs1 \citep{trujillo04}, 
22\farcs8 \citep[RC3]{devaucouleurs91}, and 25\arcsec \citep{burstein87}.
We decided to adopt the average of the above values, $R_{\rm e}$ = 
19\farcs6 $\pm$ 3\farcs7 as the effective bulge radius of NGC 5077.   
At the adopted distance of the galaxy, this value corresponds to 3.6 $\pm$
0.7 kpc. 

The stellar velocity dispersion determinations we found in the literature 
are 285 $\pm$ 28 \kms \citep[1\farcs5 $\times$ 4\arcsec aperture]{davies87}, 
252 $\pm$ 12 \kms \citep[1\farcs3 $\times$ 5\arcsec and 1\farcs6 $\times$ 
5\arcsec apertures]{carollo93}, and 263 $\pm$ 5 \kms \citep[1\farcs5 $\times$ 
1\farcs8 aperture]{caon00}. All the above values refer to the very central 
regions of the galaxy, corresponding to equivalent circular apertures 
\citep[see][]{cappellari06} from $\sim$ 1\arcsec to 1\farcs5. We then 
normalized the values of the central velocity dispersion to a circular 
aperture of radius equal to the adopted $R_{\rm e}$, following the formula 
for aperture corrections by \citet{cappellari06}. The average of these 
normalized values leads to $\sigma_{\rm R_{\rm e}}$ = 222 $\pm 15$ \kms. 

\begin{figure}
\centering
\psfig{figure=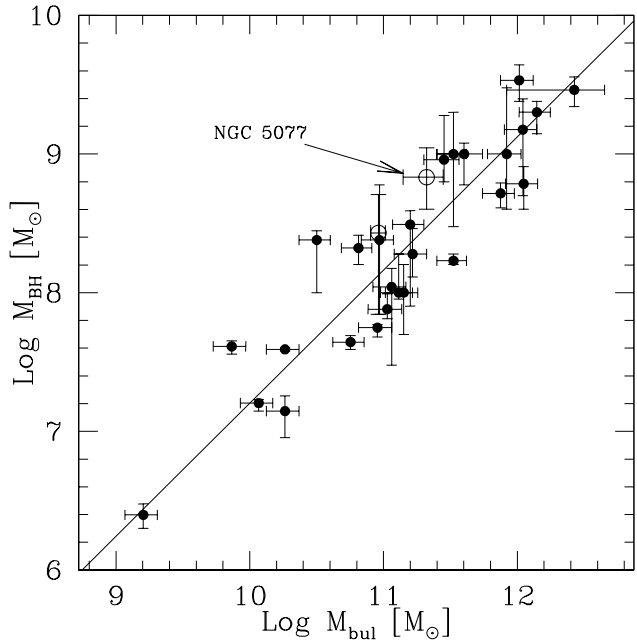,width=1.0\linewidth,angle=0}
\caption{\label{Marconi} $M_{\rm BH}$ vs bulge mass from \citet{marconi03} 
with the best fit obtained from a bisector linear regression analysis (solid
line). The positions of NGC 5077 and NGC 3998 are shown by empty circles.} 
\end{figure}

Using this value of $\sigma$ as an approximation for $\sigma_{\rm bul}$, 
we obtained the value $M_{\rm bul} = (2.1 \pm 0.7) \times 10^{11} M_{\odot}$ 
for the bulge mass. From the correlation of \citet{marconi03}, 
the expected $M_{\rm BH}$ for NGC 5077 is 3.0 $\times 10^{8} M_{\odot}$, 
in agreement within a factor $\sim$ 2.3, with our determination 
(see Fig.\ref{Marconi}).

\begin{figure}
\centering
\psfig{figure=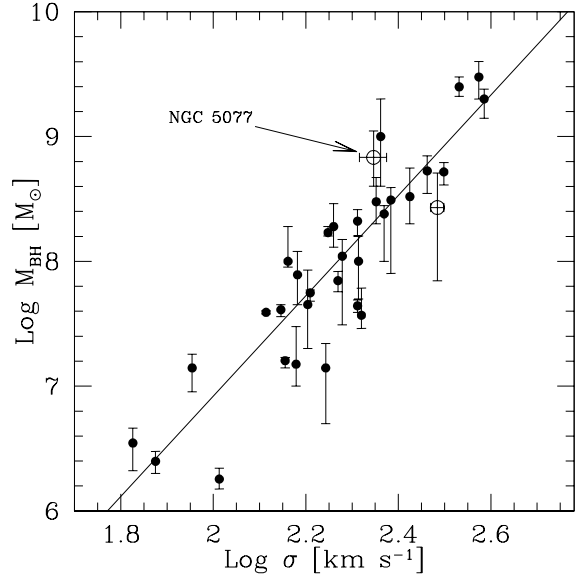,width=0.9\linewidth,angle=0}
\centering
\psfig{figure=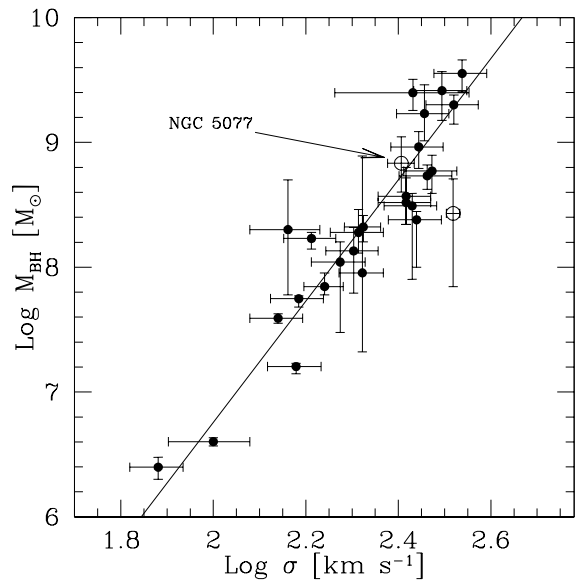,width=0.9\linewidth,angle=0}
\caption{\label{Mbh_sigma}$M_{\rm BH}$ vs central stellar velocity dispersion 
from \citet{tremaine02} ({\it upper panel}) and from \citet{ferrarese05} 
({\it lower panel}). Solid lines indicate the best fits for the two 
correlations. The positions of NGC 5077 and NGC 3998 are shown by empty 
circles.}
\end{figure}

For the correlations between BH mass and central stellar velocity 
dispersion, one obtains 2.0 $\times 10^8  M_{\odot}$ for NGC 5077 adopting 
the form by \citet{tremaine02} (see Fig.\ref{Mbh_sigma}, upper panel), a
factor 3.4 lower than our estimate. We also compared our measurement with 
the expected value from the $M_{\rm BH}-\sigma$ relation derived by 
\citet{ferrarese05}\footnote{We normalized the central velocity dispersion 
to an aperture of radius equal to 1/8 of $R_{\rm e}$ and derived 
$\sigma_{\rm R_{\rm e}/8} = 255 \pm 17$ \kms.}, 
$M_{\rm BH}$ = 5.4 $\times 10^{8}  M_{\odot}$, in full agreement with 
our estimate (see Fig.\ref{Mbh_sigma}, bottom panel).  

Conversely, the BH mass expected for NGC 5077 on the basis of its correlation
with the S{\' e}rsic concentration index \citep[from][]{trujillo04} is a 
factor 5$\sim$6 lower than our estimate when using 
the linear $M_{\rm BH} -$ S{\' e}rsic index correlation \citep{graham07} 
and a factor 3 lower when using the quadratic form they proposed. 

\begin{figure}
\centering
\psfig{figure=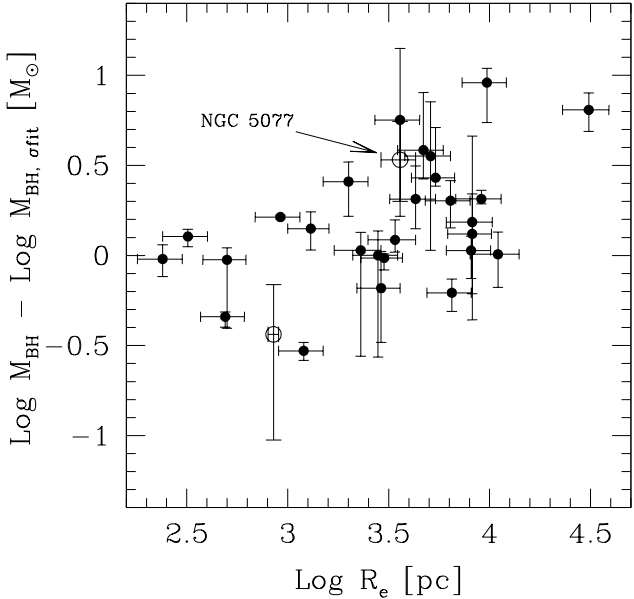,width=1.0\linewidth,angle=0}
\caption{\label{residuals} Residuals from the $M_{\rm BH}$ vs $\sigma$
correlation (in the Tremaine et al. form) reported against the bulge 
effective radius $R_{\rm e}$. The positions of NGC 5077 and NGC 3998 
are indicated by empty circles.}
\end{figure}

\citet{marconi03} noticed that $M_{\rm BH}$ is separately correlated with 
both $\sigma$ and $R_{\rm e}$. This is shown by the weak correlation between 
the residuals of the $M_{\rm BH}-\sigma$ correlation with $R_{\rm e}$ 
reproduced here in Fig. \ref{residuals} for the Tremaine at al. 
form\footnote{A similar trend is obtained using the correlation in the 
Ferrarese et al. form.}. Our determinations of the black hole mass in 
NGC 5077 and in NGC 3998 (Paper I) support this idea. Unlike NGC 3998, 
which has one of the lowest values of $R_{\rm e}$ among galaxies with 
measured $M_{\rm BH}$ (0.85 kpc) and shows a negative residual from the 
$M_{\rm BH}-\sigma$ correlation, NGC 5077 has an intermediate $R_{\rm e}$ 
value (3.6 kpc) and show a small but positive residual 
(see Fig. \ref{residuals}). In the same sense, but with a larger positive 
residual, there is the result found by \citet{Capetti05} for NGC 5252, a 
galaxy with quite a large effective radius (9.7 kpc). This indicates the 
presence of a black hole's ``fundamental plane'' in the sense that a 
combination of at least $\sigma$ and $R_{\rm e}$ drives the correlations 
between $M_{\rm BH}$ and the bulge properties. The physical implications 
of these results will be discussed in a forthcoming paper, when new direct 
BH mass measurements are presented, in order to base our discussion on a 
statistically more significative sample of $M_{\rm BH}$ determinations. 

\section{Summary and conclusions}
\label{summary}

We have presented results from a gas kinematics study in the nucleus of 
three nearby LINER galaxies: IC 989, NGC 5077, and NGC 6500 using archival 
HST/STIS spectra. Only in the case of NGC 5077 does the nuclear velocity 
curves appear to be associated with gas in regular rotation. For IC 989 our 
results indicate an inner counter-rotating gas system, while for NGC 6500 
the complex trend in the velocity curves suggests a nuclear expanding bubble.
 
We used our modeling code to fit the observed [N II]$\lambda$6583 
surface brightness distribution and velocity curve of NGC 5077. 
The dynamics of the rotating gas can be accurately reproduced by motions
in a thin disk when a compact dark mass of $M_{\rm BH} =
6.8_{-2.8}^{+4.3}\times 10^8  M_{\odot}$ is added to the stellar mass 
component. Furthermore, the black hole in NGC 5077 has a sphere of 
influence radius, $R_{\rm sph} = GM_{\rm BH}/\sigma_{\rm star}^{2}$, 
of $\sim$ 62 pc ($\simeq$ $0\farcs34$), well-resolved at the HST
resolution (2$R_{\rm sph}/R_{\rm res} \simeq$ 6.8).

For what concerns the connections of this BH mass estimate with the
properties of the host galaxy, the $M_{\rm BH}$ value for NGC 5077 is
in good agreement (within a factor of 2.3) with the $M_{\rm BH}-M_{\rm bul}$ 
correlation between BH and host bulge mass. The black hole mass predicted by 
the $M_{\rm BH}-\sigma$ correlation is a factor of 3.4 lower than our measure, 
adopting the relation found by \citet{tremaine02} and  in excellent agreement 
using the parameterization by \citet{ferrarese05}. 

This result, in conjunction with the previous results for NGC 3998 (Paper I),  
strengthens the possibility of a connection between the residuals from the 
$M_{\rm BH}-\sigma$ relation and the bulge effective radius. While NGC 3998, 
indeed, has one of the lowest values of $R_{\rm e}$ among galaxies with 
measured $M_{\rm BH}$ and shows a negative residual, NGC 5077 has a larger 
effective radius and shows a small positive residual. We also recently showed 
that the same result was found for the Seyfert galaxy NGC 5252: a larger 
effective radius corresponds in this galaxy to a still larger positive 
residual.

Apparently, only with a combination of at least $\sigma$ and $R_{\rm e}$ is 
it possible to account for the correlations between $M_{\rm BH}$ and other 
bulge properties, indicating the presence of a black hole's ``fundamental
plane''. Clearly, the number of direct black-hole mass measurements must be 
further increased, together with precise determinations of $\sigma$ and 
$R_{\rm e}$, to test these conclusions on a stronger statistical basis. In a 
forthcoming paper we will present new BH mass determinations and discuss the 
physical implications of our results. 

\begin{acknowledgements}
This publication makes use of the HyperLeda database, available at 
http://leda.univ-lyon1.fr, and of the NASA/IPAC Extragalactic Database 
(NED) operated by the Jet Propulsion Laboratory, California 
Institute of Technology, under contract with the National Aeronautics 
and Space Administration. 
\end{acknowledgements}

\end{document}